\documentclass[aps,reprint,amsmath,amssymb]{revtex4-1}
\usepackage{url}
\usepackage{float}
\usepackage{graphicx}
\usepackage{dcolumn}
\usepackage{bm}
\usepackage[portuguese]{babel}
\usepackage[utf8]{inputenc}
\usepackage[T1]{fontenc}
\begin{document}
\title{Proton Pressure: A Simplified Model}
\author{Julliana Marinho Paszko}
\email{jullianamarinho@gmail.com}
\affiliation{Centro de Ciências Naturais e Humanas, Universidade Federal do ABC (UFABC), Av. dos Estados 5001, 09210-580, Santo André, SP.}

\author{Mauro Rogério Cosentino}
\email{mauro.cosentino@ufabc.edu.br}
\affiliation{Centro de Ciências Naturais e Humanas, Universidade Federal do ABC (UFABC), Av. dos Estados 5001, 09210-580, Santo André, SP.}

\date{August 10$^{th}$, 2018}

\begin{abstract}

In this work, the authors present a simplified method for determining proton pressure, based on concepts from statistical mechanics, quantum chromodynamics - specifically asymptotic freedom of quarks - and other well-established physical concepts. The result obtained for the proton pressure was $0.3570\times10^{35}$ Pa, which is consistent with recently reported experimental data. 

\begin{description}
\item[Keywords]
proton pressure, MIT bag model, fermion gas. 
\end{description} 

\end{abstract}

\maketitle

%\tableofcontents

\section{Introduction}

Recently, the first measurement of the proton's internal pressure distribution was published \cite{Burkert}, showing it to be the highest known pressure to date, on the order of $10^{35}$ Pa. This initially surprising result can have its order of magnitude calculated from well-established concepts in statistical mechanics and quark asymptotic freedom \cite{Gross,Politzer}.

We will use a particular interpretation of asymptotic freedom by adopting a simplified version of the MIT Bag Model \cite{Chodos}. In this model, the proton is viewed as a spherical cavity in which constituent quarks are confined but free to move. The hypothesis of quark mobility is based on the fact that at short distances, QCD interaction is weak, so quarks behave almost as free particles. However, as the distance between quarks increases, the interaction rapidly strengthens, leading to confinement. Thus, an approximation would be that within the proton volume, quarks are free particles, but the proton boundaries act as a potential barrier, forcing confinement.

Next, using Fermi-Dirac statistics, we will assume that confined quarks behave as a Fermi gas and derive an expression for calculating proton pressure. We will present the results and discuss the assumed conditions.

\section{Proton Modeling: Asymptotic Freedom and the MIT Bag Model}

According to the Standard Model, a proton is a baryon composed of 3 quarks - two \textit{up} and one \textit{down}. The predominant force between quarks is the strong force, described by Quantum Chromodynamics (QCD). Both quarks and gluons, the mediating particles of the strong force, are assigned charges called \textit{color}.

To date, no free particle with color charge has been detected, leading to the belief that they are confined within baryons and mesons.  

The strong force is powerful enough to keep quarks confined, but at short distances, the phenomenon of asymptotic freedom occurs. The closer the quarks are, the weaker the coupling becomes, and consequently, quarks interact much more weakly. 
\begin{center}
\begin{figure}[!ht]
\centering
\includegraphics[scale=.60]{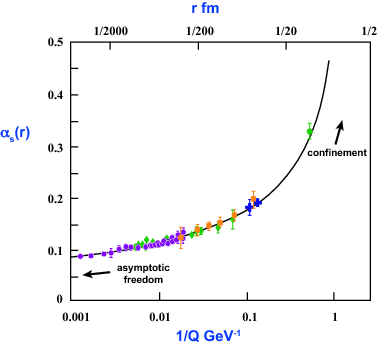}
\caption{\label{fig:asymptoticfreedom}Plot of the coupling constant $\alpha_s$ as a function of distance $r$ \cite{Pennington}.}
\end{figure}
\end{center}

The QCD coupling constant $\alpha_{s}$, shown in Fig.~\ref{fig:asymptoticfreedom}, varies with distance $r$, where $r=1/Q$, and $Q$ is the momentum scale felt in the interaction, displayed on a logarithmic scale. At short distances, the coupling is weak - asymptotic freedom. At large distances ($r>1$ fm), the coupling becomes strong, leading to confinement.

Thus, we can consider the approximation that the quarks constituting the proton move as free particles.

The asymptotic freedom regime allows calculations using perturbation theory, and its discovery enabled QCD theory to develop similarly to QED. For the confinement regime, perturbation theory is not applicable, making its theoretical demonstration difficult.

Given this difficulty, several confinement models emerged, including the MIT Bag Model. In this model, the fields describing quarks and gluons are restricted to a specific region called a "bag". A surface is specified, and it is required that color current does not pass through it. Since both quarks and gluons carry color charge, neither can cross this barrier.

Therefore, we will use this concept to simplify the proton as a spherical bag containing a gas of free quarks. 

\section{Ideal Fermi Gas}

We know that quarks are particles governed by Fermi-Dirac statistics, as they have half-integer spin, specifically spin $1/2$. Thus, in our model, we assume the proton consists of a non-relativistic fermion gas. To describe it, we use the grand canonical ensemble formalism \cite{Pathria}. Its partition function $\mathcal{Z}$ is given by 
\begin{equation}\label{eq:grandcanonical}
    \mathcal{Z}(V,T,z)= \sum_{N=0}^{\infty} z^N Q_N,
\end{equation}
where $V$ is volume, $T$ is temperature, and
\begin{equation}
Q_N=\sum_{E} e^{-\beta E}\label{canonicalpartition}
\end{equation}
is the canonical partition function, obtained by summing over all energy values of a system with N particles. Here $z=\exp({\beta \mu})$ is the fugacity, with $\displaystyle \beta=\frac{1}{kT}$, and $\mu$ is the chemical potential.

Since we are considering free particles, the total energy $E$ is the sum of the energies of each particle, which can be written as 
\begin{equation}
    E=\sum_{i} n_i\epsilon_i, \label{energysum}
\end{equation}
where $\epsilon_i$ is the energy of the i-th level (assuming quantized energies), and $n_i$ is the number of particles in that level. 

Thus, the total number $N$ of particles must satisfy
\begin{equation}
    \sum_i n_i=N. \label{N}
\end{equation}

Substituting (\ref{energysum}) into (\ref{canonicalpartition}), we get
\begin{equation}\label{partition}
    Q_N=\sum_{\{n_i\}}{}^{'}\exp\left(-\beta\sum_in_i\epsilon_i\right),
\end{equation}
where the primed summation $\displaystyle\sum_{\{n_i\}}{}^{'}$ means summing over all configurations $\{n_i\}=\{n_1,n_2,n_3, \ldots \}$ of occupation numbers satisfying condition (\ref{N}), i.e., keeping the total number of particles equal to N.

Now substituting (\ref{partition}) into (\ref{eq:grandcanonical}), we have
\begin{align*}
    \mathcal{Z}(V,T,z)&= \sum_{N=0}^{\infty}z^N\sum_{\{n_i\}}{}^{'}\exp\left(-\beta\sum_in_i\epsilon_i\right) \\
    &=\sum_{N=0}^{\infty}\sum_{\{n_i\}}{}^{'}z^N\prod_i\left(e^{-\beta\epsilon_i}\right)^{n_i}
\end{align*}
where the exponential summation becomes a product. Similarly,
 \begin{equation*}
     z^N=z^{\sum_i n_i}=\prod_i z^{n_i}.
 \end{equation*}
Finally,
\begin{equation*}
    \mathcal{Z}(V,T,z)=\sum_{N=0}^{\infty}\sum_{\{n_i\}}{}^{'}\prod_{i}\left(ze^{-\beta\epsilon_i}\right)^{n_i}.
\end{equation*}
 The two summations in the previous expression are equivalent to summing over all values of $n_i$ independently:
 \begin{align}
     \mathcal{Z}(V,T,z)&=\sum_{n_0,n_1,...}\left(ze^{-\beta\epsilon_0}\right)^{n_0}\left(ze^{-\beta\epsilon_1}\right)^{n_1} ...= \nonumber \\
     &=\left[\sum_{n_0}\left(ze^{-\beta\epsilon_o}\right)^{n_0}\right]\left[\sum_{n_1}\left(ze^{-\beta\epsilon_1}\right)^{n_1}\right]... \label{sum_grandpartitionfunction}
 \end{align}
 
Since we are dealing with fermions, for each energy level the occupation number can only have two possible values: either the level is occupied by one fermion, or unoccupied:
\begin{equation*}
    n_i=\begin{cases}
    0, \quad\text{unoccupied level} \\
    1, \quad\text{level occupied by one fermion.}
    \end{cases}
\end{equation*}
 Therefore, for the sums in Eq.~(\ref{sum_grandpartitionfunction}), we have
\begin{align*}
    \mathcal{Z}(V,T,z)&=\left(1+ze^{-\beta\epsilon_0}\right)\left(1+ze^{-\beta\epsilon_1}\right)\ldots \\
    &=\prod_i\left(1+ze^{-\beta\epsilon_i}\right).
\end{align*}

The connection between the grand canonical ensemble and thermodynamics is given by $PV=kT \ln \mathcal{Z}$ (see Ref.~\cite{Pathria}), so
\begin{equation}\label{eq:idealfermigas}
    \frac{PV}{kT}=\ln\mathcal{Z}(V,T,z)=\sum_i \ln(1+ze^{-\beta\epsilon_i}),
\end{equation}
where we used the logarithmic property that converts a product into a sum.

In the thermodynamic limit ($V\to\infty$), energy levels form a continuum. Thus, we replace the summation with phase space integrals
\begin{equation*}
    \sum_i \cdots \rightarrow\frac{1}{h^3}\int \cdots d^3qd^3p.
\end{equation*}

Integrating $d^3q$ gives volume $V$, and $d^3p$ can be replaced by $4\pi p^2dp$ using spherical coordinates:
\begin{align}
    \frac{1}{h^3}\int \cdots d^3qd^3p&=\frac{V}{h^3}\int \cdots d^3p \nonumber \\
    &=\frac{V}{h^3}\int \cdots 4\pi p^2dp. \label{integralconvertion}
\end{align}
    
The non-relativistic momentum $p$ is
\begin{equation}
    p=(2m\epsilon)^{1/2} \label{p},
\end{equation}
and $dp$ is
\begin{equation}
    dp=\frac{1}{2}(2m)^{1/2}\epsilon^{-1/2}d\epsilon. \label{dp}
\end{equation}

Substituting Eqs. (\ref{p}) and (\ref{dp}) into (\ref{integralconvertion}):
 \begin{align*}
     \frac{V}{h^3}\int \cdots 4\pi p^2dp=&\frac{V}{h^3}4\pi\int \cdots (2m\epsilon)\frac{1}{2}(2m)^{1/2}\epsilon^{-1/2}d\epsilon \\
     =&\frac{V}{h^3}4\pi\frac{1}{2}(2m)^{1/2}2m\int \cdots (\epsilon)\epsilon^{-1/2}d\epsilon \\
     =&\frac{V}{h^3}2\pi(2m)^{3/2}\int \cdots \epsilon^{1/2}d\epsilon.
 \end{align*}

Applying this summation-to-integral replacement to Eq.~(\ref{eq:idealfermigas}):
\begin{equation*}
    \frac{PV}{kT}=2\pi \frac{V}{h^3}(2m)^{3/2}\int_0^{\infty}\ln(1+ze^{-\beta\epsilon})\epsilon^{1/2}d\epsilon.
\end{equation*}

Changing variables to $x=\beta\epsilon$ ($\displaystyle\epsilon=\frac{x}{\beta}$, $\displaystyle d\epsilon=\frac{1}{\beta}dx$):
\begin{align*}
    \frac{PV}{kT}=&2\pi \frac{V}{h^3}(2m)^{3/2}\int_0^{\infty}\ln(1+ze^{-x})\frac{x^{1/2}}{\beta^{1/2}}\frac{1}{\beta}dx \\
    =&2\pi \frac{V}{h^3}(2m)^{3/2}\frac{1}{\beta^{3/2}}\int_0^{\infty}ln(1+ze^{-x})x^{1/2}dx.
\end{align*}

Integrating by parts with $\displaystyle u=\ln(1+ze^{-x})$ and $dv=x^{1/2}dx$ ($\displaystyle du=-\frac{ze^{-x}}{1+ze^{-x}}dx$, $\displaystyle v=\frac{2}{3}x^{3/2}$):
\begin{align*}
    \int_{0}^{\infty}\ln (1+ze^{-x})dx&=uv\biggr\rvert_0^{\infty} - \int_{0}^{\infty} v du  \\
    &=\ln(1+ze^{-x})\frac{2}{3}x^{3/2}\biggr\rvert_0^{\infty}  \\
    &\qquad\quad+\frac{2}{3}z \int_{0}^{\infty}\frac{x^{3/2}e^{-x}}{1+ze^{-x}}dx  \\
    &= \frac{2z}{3}\int_{0}^{\infty}\frac{x^{3/2}}{e^x+z}dx  \\
    &=\frac{2}{3}\int_{0}^{\infty}\frac{x^{3/2}}{z^{-1}e^x+1}dx.
\end{align*}

Therefore,
\begin{align*}
    \frac{PV}{kT}=&2\pi \frac{V}{h^3}\frac{(2m)^{3/2}}{\beta^{3/2}}\frac{2}{3}\int_{0}^{\infty}\frac{x^{3/2}}{z^{-1}e^x+1}dx \\
    =&V\frac{4}{3\sqrt{\pi}}\left[\frac{(2\pi mkT)^{3/2}}{h^3}\right]\int_{0}^{\infty}\frac{x^{3/2}}{z^{-1}e^x+1}dx  \\
    =&V\frac{4}{3\sqrt{\pi}}\frac{1}{\lambda^3}\int_{0}^{\infty}\frac{x^{3/2}}{z^{-1}e^x+1}dx,
\end{align*}
where we substituted the thermal de Broglie wavelength $\displaystyle\lambda=\frac{h}{\sqrt{2\pi mkT}}$.

The integral corresponds to the Fermi-Dirac function \cite{Pathria}:
\begin{equation*}
    f_\nu(z)=\frac{1}{\Gamma(\nu)}\int_0^{\infty}\frac{x^{\nu-1}dx}{z^{-1}e^x+1},
\end{equation*}
where $\displaystyle\Gamma(\nu)=\int_0^{\infty}x^{\nu-1}e^{-x}dx$ is the gamma function. Finally:
\begin{equation*}
    \frac{PV}{kT}=\frac{V}{\lambda^3}f_{5/2}(z) \label{eq:pvkt}.
\end{equation*}

We now calculate the internal energy:
\begin{align*}
    U&=-\left.\frac{\partial}{\partial\beta}\ln\mathcal{Z}\right|_{z,V}  \\ 
    &=kT^2\frac{\partial}{\partial T}\left(\frac{PV}{kT}\right)_{z,V} \nonumber \\ 
    &=kT^2\frac{\partial}{\partial T}\left(\frac{V}{\lambda^3}f_{5/2}(z)\right)  \\
    &=\frac{3}{2}kT\left(\frac{V}{\lambda^3}f_{5/2}(z)\right)  \\
    &=\frac{3}{2}kT\left(\frac{PV}{kT}\right) \nonumber \\
    &=\frac{3}{2}PV.
\end{align*}
Equivalently:
\begin{equation}
    P=\frac{2}{3}\frac{U}{V} \label{eq:pressure}.
\end{equation}

Note that this result is identical to that of a classical monoatomic ideal gas.

\section{Pressure Determination}

In this model, the internal energy can be considered as the proton's rest energy:
\begin{equation*}
    U = m_p c^2,
\end{equation*}
so by Eq.~(\ref{eq:pressure}), the proton's internal pressure is:
\begin{equation*}
    P=\frac{2}{3}\frac{m_{p}c^2}{V}.
\end{equation*}

For the volume, we assume quarks are confined to a sphere with radius equal to the proton radius:
\begin{align}
    P=&\frac{2}{3}\cdot\frac{3}{4}\frac{m_pc^2}{\pi r_p^3} \nonumber \\
    =&\frac{1}{2}\frac{m_pc^2}{\pi r_p^3}.
\end{align}

Using the proton radius value from Ref.~\cite{PDG} (CODATA 2014) and the proton mass from the same reference:
\begin{equation}
  P=0.3570\times 10^{35} \text{ Pa}.  
\end{equation}

\section{Results and Conclusion}

We discussed a simplified model for the proton where, due to quark confinement by the strong force, we consider them trapped in a volume defined by a spherical surface. Due to asymptotic freedom at short distances, we treat them as free particles, approximating the proton as a quantum gas of fermions at constant volume.

This model allowed calculation of proton pressure, achieving the same order of magnitude as recent experimental results \cite{Burkert}. 

The Fermi-Dirac statistics requires the thermodynamic limit ($V\to\infty$). This is justified because the volume occupied by quarks is much smaller than the proton volume, consistent with the point-like nature of quarks.

We assumed a non-relativistic Fermi gas. While quarks could reach high speeds within the proton, the ultra-relativistic case would yield $\displaystyle P=\frac{1}{3}\frac{U}{V}$ \cite{Greiner} (losing a factor of 2 compared to Eq.~\ref{eq:pressure}), not changing the final order of magnitude.  

This result is significant because basic physical concepts yield good agreement with recent experimental data. Achieving the same order of magnitude validates the approximations and demonstrates that important information can be extracted from simple models.

\bibliography{references}{}
\bibliographystyle{unsrt}

\end{document}